\documentclass[9pt,twocolumn,twoside]{opticajnl}
\journal{opticajournal} 

\setboolean{shortarticle}{true}

\usepackage{lineno}
\usepackage{amsmath}

\usepackage{siunitx}
\usepackage{physics}
\newcommand{\cuprite}{Cu$_\mathrm{2}$O}

\title{Giant microwave-optical Kerr nonlinearity via Rydberg excitons in cuprous oxide}

\author[1,$\dagger$]{Jon D. Pritchett}
\author[1,$\dagger$]{Liam A. P. Gallagher}
\author[1]{Alistair Brewin}
\author[1]{Horatio Q. X. Wong}
\author[2]{Wolfgang Langbein}
\author[2]{Stephen A. Lynch}
\author[1]{C. Stuart Adams}
\author[1,*]{Matthew P. A. Jones}

\affil[1]{Department of Physics, Durham University, Durham DH1 3LE, United Kingdom}
\affil[2]{School of Physics and Astronomy, Cardiff University, Cardiff CF24 3AA, United Kingdom}

\affil[$\dagger$]{These authors contributed equally to this work.}
\affil[*]{m.p.a.jones@durham.ac.uk}

\begin{abstract}
Microwave-optical conversion is key to future networks of quantum devices, such as those based on superconducting technology. Conversion at the single quantum level requires strong nonlinearity, high bandwidth, and compatibility with a millikelvin environment. A large nonlinearity is observed in Rydberg atoms,  but combining atomic gases with dilution refrigerators is technically challenging. Here we demonstrate that a strong microwave-optical nonlinearity in a cryogenic, solid-state system by exploiting Rydberg states of excitons in \cuprite. We measure a microwave-optical cross-Kerr coefficient of $B_0 = 0.022 \pm 0.008 $ m V$^{-2}$ at 4~K, which is several orders of magnitude larger than other solid-state systems. Our results highlight the potential of Rydberg excitons for nonlinear optics, and form the basis for a  microwave-optical frequency converter based on \cuprite.
\end{abstract}

\setboolean{displaycopyright}{false} 

\begin{document}

\maketitle

Superconducting microwave devices play a key role in quantum computation \cite{Kjaergaard2020}. To eliminate thermal noise, these devices must be cooled to $T\approx 10$ mK, a requirement which makes direct quantum networking impractical over distances larger than $\sim1$~m~\cite{Storz2023}. Microwave-optical (MO) conversion is therefore a critical enabling technology \cite{Lauk2020,Lambert2020,Han2021}, with current approaches including electro-optics~\cite{Wang2022}, rare-earth ions~\cite{Bartholomew2020}, optomechanical systems~\cite{Honl2022} and quantum dot molecules~\cite{Tsuchimoto2022}. Very strong nonlinearity can be achieved by exploiting the large microwave dipole moment associated with highly excited atomic Rydberg states~\cite{Han2018}, leading to the largest observed cross-Kerr effect of any material~\cite{Mohapatra2008}. However interfacing Rydberg atoms with planar superconducting quantum devices in a millikelvin environment is an outstanding challenge, and coupling has so far been achieved only at much higher temperature~\cite{Hogan2012,Hermann2014,Morgan2020,Kaiser2022}.

\begin{figure*}
    \centering
    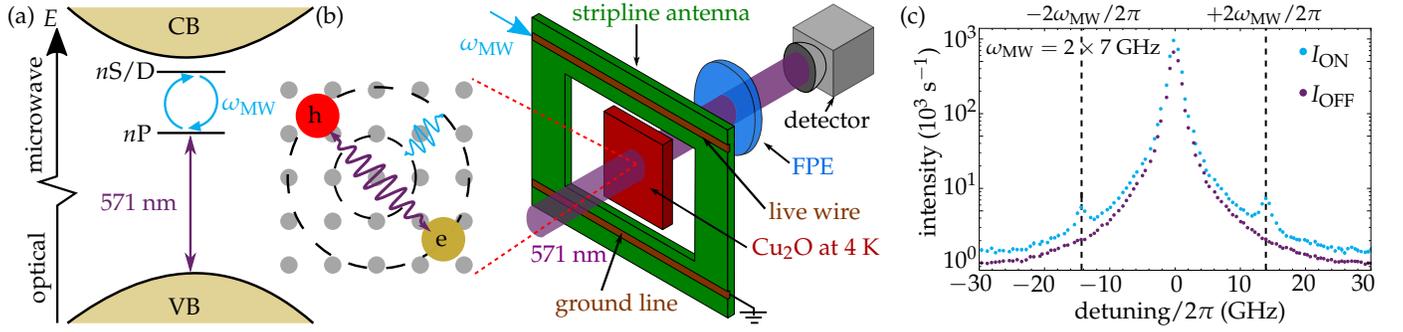
    \caption{
    (a) Excitons are created in state $n$P via laser excitation at $\lambda \approx 571$~nm. A microwave field at frequency $\omega_\mathrm{MW}$ couples odd- and even-parity exciton states. The exciton wavefunction spans many lattice sites giving rise to a large dipole moment. (b) Excitons are created in a thin slab of \cuprite\ located between the conductors of a microwave stripline.  The transmitted light is filtered with a Fabry-P\'erot etalon (FPE) and sent to a detector. (c) Example spectrum of the transmitted light for $E=h c/ \lambda=2.170127$~eV and $\mathcal{E}_\mathrm{MW}=170$~V/m, showing the appearance of sidebands at $\pm 2\omega_\mathrm{MW}/(2\pi)$ when microwaves are applied  ($I_\mathrm{ON}$).}
    \label{fig:setup}
\end{figure*}

In this Letter we combine the advantages of atomic Rydberg states (giant nonlinearity) and the solid state (milliKelvin compatibility~\cite{Heckotter2020,versteegh2021}) by using Rydberg states of excitons in the bulk semiconductor \cuprite~\cite{Kazimierczuk2014} (Fig.~\ref{fig:setup}). Excitons are optically excited bound states of an electron and a hole, with an internal structure that resembles a hydrogen atom. In \cuprite, the resulting Rydberg series of excitonic states has been measured up to principal quantum number $n=30$ at $T\approx40$~mK~\cite{versteegh2021}. Transitions between neighbouring states of opposite parity are located in the microwave spectral region. The associated dipole moment scales as $d\propto n^2$,  reaching 180~$e$~nm at $n=11$, which is more than thirty times larger than in quantum dot molecules~\cite{Tsuchimoto2022}. Thus  Rydberg excitons in \cuprite\ provide a unique platform for MO coupling in the solid-state~\cite{Ziemkiewicz2023}, with recent experiments demonstrating microwave control of the linear and nonlinear optical response~\cite{Gallagher2022}. Here we extend this work to a measurement of the optical phase shift induced by the presence of a microwave field, characterised by the microwave-optical Kerr coefficient $B_0$ 
\begin{equation}
    B_0 = \frac{\Delta\phi}{2 \pi L \left|\mathcal{E}_\mathrm{MW}\right|^2},
    \label{eq:kerr}
\end{equation}
where $\Delta \phi$ is the optical phase shift induced by the microwave field,
$L$ is the length of the material and $\mathcal{E}_\mathrm{MW}$ is the amplitude of the applied microwave field. 

A schematic of the experiment is shown in Fig.~\ref{fig:setup}. A thin slab of \cuprite\ (55$\pm$10~\si{\micro\meter}) is mounted between the conductors of a microwave stripline and cooled to $T=4$~K. Excitons in a quantum state $n$P (where P denotes orbital angular momentum $l=1$) were excited from the ground state (valence band) using a single-frequency laser at wavelength $\lambda \approx 571 $~nm. A microwave field at angular frequency $\omega_\mathrm{MW} = 2\pi\times 7$~GHz couples $n$P states to nearby states of opposite parity (e.g. $n^\prime$S, $n^\prime$D, where S and D indicate $l=0$ and $l=2$). The spectrum of the transmitted laser light was resolved using a scanning Fabry-Perot etalon. Further experimental details are provided in Appendix~\ref{App:experiment}.  Applying the microwave field leads to a change in transmission at the laser frequency $\omega_\mathrm{L}= 2\pi c / \lambda$, and the appearance of sidebands at frequencies $\omega_\mathrm{L}\pm 2\omega_{\mathrm{MW}}$ on the light transmitted through the sample (Fig.~\ref{fig:setup}(c)) that are indicative of a cross-Kerr nonlinearity. 

\begin{figure}
    \centering
    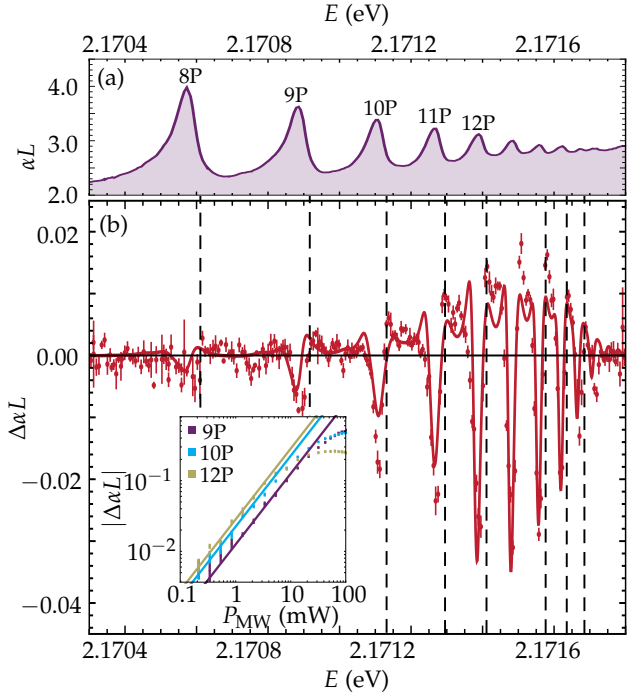
    \caption{(a) Rydberg series of exciton absorption lines. (b) Measured (red points) and predicted (solid line) change in absorption, $\Delta\alpha L$ at $\mathcal{E}_\mathrm{MW}=38$~V/m. Dashed lines indicate the  zero crossings used to extract the Kerr coefficient. Inset shows $\Delta \alpha$ versus microwave power $P_\mathrm{MW}$ at the $n=9,10,12$P resonances. Solid lines are fits to the linear region.}
    \label{fig:change_in_abs}
\end{figure}

The optical absorption spectrum close to the bandgap is shown in Fig.~\ref{fig:change_in_abs}(a). Peaks corresponding to $n$P excitonic states are visible up to $n=16$, limited by a combination of temperature and sample quality. As \cuprite\ is centrosymmetric, the MO coupling absent from the ground-state symmetry but is created by the Rydberg excitons, as shown in Fig.~\ref{fig:change_in_abs}(b), which plots the spectrum of the microwave-induced change in absorption $\Delta \alpha$. Each exciton resonance shows decreased absorption on-resonance, and increased absorption on each side at the location of the $n$S and $n$D states. In contrast to Rydberg atoms, a response is observed over a wide range of $n$, even for a single microwave frequency. This is due to non-radiative broadening of the exciton lines, which means that the microwave field is near-resonant with many transitions simultaneously. The same effect results in a broad microwave frequency dependence (see Appendix \ref{App:experiment} and \cite{Gallagher2022}). $|\Delta\alpha|$ increases with $n$ up to $n=13$ before decreasing towards the band edge. As shown in the inset of Fig.~\ref{fig:change_in_abs}(b) $|\Delta \alpha|$ increases linearly with the microwave power as expected for a Kerr nonlinearity, before saturating at a value which depends on $n$. Saturation occurs due to multi-photon processes, and is accompanied by the appearance of higher-order sidebands. For the remainder of this paper we consider only the linear regime.

The MO Kerr nonlinearity can be described in terms of the dielectric polarization at the laser frequency
\[
\mathcal{P}(\omega_\mathrm{L}) = \epsilon_0\left( \chi^{(1)}(\omega_\mathrm{L})\mathcal{E}_\mathrm{L} +  \chi^{(3)}(\omega_\mathrm{L})\mathcal{E}_\mathrm{L}\mathcal{E}_\mathrm{MW}\mathcal{E}_\mathrm{MW}+\ldots\right),
\label{eq:polarisation}
\]
where $\chi^{(1)}$and $\chi^{(3)}$ are the linear and Kerr nonlinear susceptibilities respectively. Experimentally we observe that the nonlinearity depends very weakly on the laser and microwave polarizations, and so a scalar treatment is used. We obtain an expression for $\chi^{(3)}$ by treating the exciton states as hydrogen-like~\cite{Gallagher2022}. $\Delta \alpha$ is related to the imaginary part of the susceptibility
\[\frac{\Delta \alpha c}{\omega_\mathrm{L}{\left|\mathcal{E}_\mathrm{MW}\right|^2}} = \mathrm{Im}\left(\chi^{(3)}(\omega_\mathrm{L})\right) =\mathrm{Im}\left(\sum_{n,n',l',n'',\pm} \chi^{(3)}_{n\mathrm{P}n'l',n''\mathrm{P}}(\omega_\mathrm{L})\right).
\]
Each term in the sum is given by
\begin{equation}
\begin{split}
    &\chi^{(3)}_{n\mathrm{P}n'l'n''\mathrm{P}}(\omega_\mathrm{L}) =\\ &\frac{N_\mathrm{d}}{8\epsilon_0 \hbar^3} \frac{D^{\mathrm{VB}\rightarrow n\mathrm{P}} d^{n\mathrm{P}\rightarrow n'l'}d^{n'l'\rightarrow n''\mathrm{P}} D^{n''\mathrm{P}\rightarrow\mathrm{VB}}}{(\delta_{n\mathrm{P}}-i\Gamma_{n\mathrm{P}})(\delta_{n'l'}^{\pm\omega_\mathrm{MW}}-i\Gamma_{n'l'})(\delta_{n''\mathrm{P}}-i\Gamma_{n''\mathrm{P}})},
\end{split}
\label{eq:chi3carrier}
\end{equation}
where $N_\mathrm{d}$ is the effective density of exciton states, $D$ are effective matrix elements for transitions from the valence band to the P states, $d$ are dipole matrix elements for transitions between exciton states, $\Gamma$ are the exciton linewidths and $\delta$ are detunings, given by $\delta_{n\mathrm{P}}=\omega_{n\mathrm{P}} -\omega_\mathrm{L}$ and $\delta^{\pm\omega_\mathrm{MW}}_{n'l'}=\omega_{n'l'} -(\omega_\mathrm{L} \pm \omega_\mathrm{MW})$. Crucially all of these parameters are either known theoretically or can be derived from analysis of data without a microwave field applied (see Appendix~\ref{App:parameters}) enabling the use of equation~\ref{eq:chi3carrier} for quantitative predictions of the nonlinear response.

The predicted $\Delta \alpha$
is compared to the data in Fig.~\ref{fig:change_in_abs}(b). The only adjustable parameter is the amplitude of the microwave electric field $\mathcal{E}_\mathrm{MW}$. The prediction is in excellent agreement with the data over the full range of $n$. Similar levels of agreement are observed up to field amplitudes of 80~V/m, above which saturation occurs. From these fits we obtain a calibration between electric field amplitude and applied microwave power of $43 \pm 3$~(V/m)/mW$^{1/2}$, in reasonable agreement with a value of 70~(V/m)/mW$^{1/2}$ obtained from a finite-element simulation of the antenna that excludes connection losses~\cite{Gallagher2022}. 

\begin{figure}
    \centering
    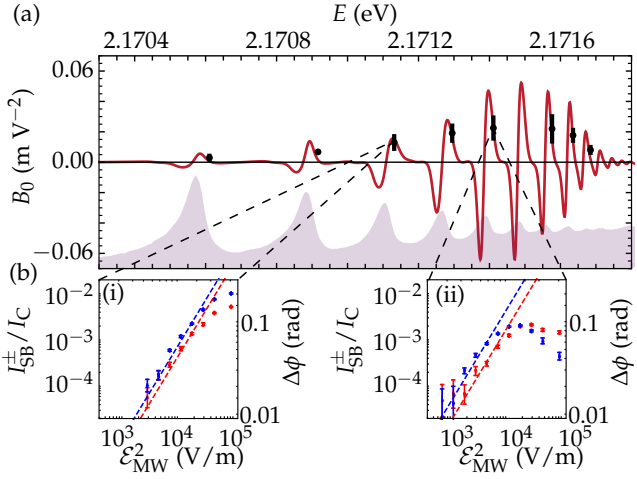
    \caption{(a) Predicted (solid line) and measured (points) Kerr coefficient vs excitation energy $E$. Shaded background shows exciton spectrum for reference. (b)  $I_\mathrm{SB}/I_\mathrm{C}$ vs $\mathcal{E}_\mathrm{MW}$ for positive frequency (blue) and negative frequency (red) sidebands. Right-hand axis gives the corresponding phase shift.}
    \label{fig:kerr}
\end{figure}

Once the microwave field amplitude is known, the real part of the susceptibility can be used to make a quantitative prediction of the microwave-optical Kerr coefficient, since 
\[
    B_0 = \frac{\omega_\mathrm{L}}{4\eta\pi c}\mathrm{Re}\left( \chi^{(3)}(\omega_\mathrm{L})\right),
\]
where $\eta=2.8$ is the optical refractive index. Predicted values of $B_0$ are shown in Fig.~\ref{fig:kerr}(a). 

To independently measure $B_0$ we make use of the zero-crossings in Fig.~\ref{fig:change_in_abs}. At these points $\Delta \alpha =0$, the imaginary part of the susceptibility is zero, and sidebands are generated via phase modulation only. By using the conventional expansion of a phase-modulated wave in terms of Bessel functions, it can be shown that the ratio of the intensity of the sidebands to the carrier is $I_\mathrm{SB}/I_\mathrm{C} = \left|J_1(\Delta \phi)\right|^2$ where $J_1$ is the first-order Bessel function of the first kind. Thus the microwave-induced phase shift $\Delta \phi$ can be directly extracted from spectra like Fig.~\ref{fig:setup}(c). 

The variation of $\Delta \phi$ with $|\mathcal{E}_\mathrm{MW}|^2$ is shown for two zero-crossings in Fig.~\ref{fig:kerr}(b).  As expected the phase shift increases linearly in the Kerr regime, before saturating. The Kerr coefficient is obtained from the gradient in the linear regime using Eq.~\ref{eq:kerr} and the field amplitude calibration from the absorption data (Fig.~\ref{fig:change_in_abs}). 

As shown in Fig.~\ref{fig:kerr}(a), the measured Kerr coefficient is in agreement with the prediction from Eq.~\ref{eq:chi3carrier}. We emphasise that the phase shift measurement does not depend on the model for $\chi^{(3)}$ (Eq.~\ref{eq:chi3carrier}) and its input parameters; these affect the measured value of $B_0$ only through the field calibration. The dominant experimental uncertainties come from the electric field calibration and the thickness of the sample. The analysis also assumes that $\chi^{(3)} (\omega_\mathrm{L} \pm 2 \omega_\mathrm{MW}) = \chi^{(3)} (\omega_\mathrm{L})$, which is not strictly true due to the strong energy dependence near resonance. An indication that this assumption does not fully hold is the asymmetry in the red and blue sideband amplitude in Fig.~\ref{fig:kerr}. We therefore include the difference between the red and blue values as part of the uncertainty in $B_0$ and present the average value. Analytic expressions for $\chi^{(3)} (\omega_\mathrm{L}\pm 2 \omega_\mathrm{MW})$ are provided in Appendix~\ref{App:sideband_chi}. 

\begin{table}
    \centering
  \begin{tabular}{|c|c|c|c|}\hline
     Material & $B_0$ (m\,V$^{-2}$)  & $\alpha$ (m$^{-1}$) & $\mathcal{F}$ (m$^2$\,V$^{-2}$)\\ \hline
     glass & $10^{-14}$~\cite{weber2002}& $10^{-1}$~\cite{weber2002} &  $10^{-13}$\\
     nitrobenzene  & $10^{-12}$~\cite{weber2002}&$10^{-1}$~\cite{Cabrera2008}&$10^{-11}$\\
    PMN-PT & $10^{-7}$~\cite{Ruan2010} & $10^{2}$~\cite{Ruan2010} & $10^{-9}$ \\
     GO-LC & $10^{-6}$~\cite{Shen2014} & $10^{3}$~\cite{Pham2011} & $10^{-9}$ \\
      Rb vapour & $10^{-6}$~\cite{Mohapatra2008} & $10^1$~\cite{Mohapatra2007} & $10^{-7}$\\
      \cuprite & $10^{-2}$&$10^{4}$ & $10^{-6}$\\ \hline
   \end{tabular}
    \caption{Comparison of Kerr ($B_0$) and absorption ($\alpha$) coefficients, and their ratio $\mathcal{F}=B_0/\alpha$ for various materials. For nitrobenzene, glass and \cuprite\  $\lambda=570$~nm; for PMN-PT (a transparent ceramic) and GO-LC (graphene oxide liquid crystals) $\lambda=633$~nm;  Rb vapour $\lambda=$ 780~nm. }
    \label{tab:comparison}
\end{table}

Our highest measured value of $B_0 = 0.022 \pm 0.008$ V m$^{-2}$ at $n=12$ is compared to other low-frequency (DC) Kerr coefficients in Table~\ref{tab:comparison}. Our measured Kerr coefficient is extremely large, due to a combination of the underlying resonant nature of the nonlinearity and the large dipole moment. The other values in Table~\ref{tab:comparison} were measured far from resonance - a key feature of \cuprite\ is that a much larger but still broadband response can be obtained via multiple resonances.   \cuprite\ also has the highest absorption coefficient among the materials in Table~\ref{tab:comparison}. An alternative figure of merit is the phase shift per unit absorption length and microwave intensity, described by the ratio $\mathcal{F}=B_0/\alpha$ where $\alpha$ is the linear absorption coefficient. Here, \cuprite\  has the largest $\mathcal{F}$, with only Rydberg atoms in Rb vapour being comparable, highlighting the importance of Rydberg physics in the solid-state.

Despite the large values of $B_0$ and $\mathcal{F}$, the maximum phase shift remained limited to around 0.1 radians as shown in Fig.~\ref{fig:change_in_abs}(b). This is due to a combination of the saturation of the phase shift at high electric field and the high background absorption, which together limit the maximum phase shift that can be measured. The saturation is a consequence of the extremely strong nonlinearity, and corresponds to the emergence of higher order terms (e.g. $\chi^{(5)}$) in the susceptibility. In fact the experiment enters the ultra-strong driving regime, where the coupling strength $\Omega_\mathrm{MW} = d \mathcal{E}_\mathrm{MW}/\hbar \gg \omega_\mathrm{MW},\Gamma$. Microwave-optical conversion still occurs in this regime, but it is no longer given by a simple Kerr effect.
A reduction in absorption would enable the use of thicker samples, resulting in larger phase shifts. The absorption is dominated by phonon-assisted processes that do not involve Rydberg states, and which are unaffected by the microwave field~\cite{Baumeister1961}. At $n=11$, 80\% of the absorption coefficient is due to the background. As in atomic Rydberg gases, the background may be reduced by nonlinear spectroscopy techniques such as second harmonic generation~\cite{Mund2018,Rogers2022} or electromagnetically induced transparency \cite{Mohapatra2007,Walther2020}, or by exploiting Rydberg exciton-polaritons \cite{Orfanakis2022}.

In conclusion, we have demonstrated that the strong microwave-optical nonlinearity observed in atomic gases can be realised in a cryogenic, solid-state setting using Rydberg excitons in \cuprite. The experimentally measured Kerr coefficient  $B_0 = 0.022\pm 0.008$ m\,V$^{-2}$ is 4 orders of magnitude larger than other solid-state platforms, with only Rydberg atoms being comparable. The observations are in agreement with a model based on the conventional hydrogen-like theory of Wannier-Mott excitons, enabling the quantitative design of future devices that exploit the giant Kerr effect, additionally we highlight the effect is visible at relatively low $n$ meaning that synthetic material could be utilised~\cite{Lynchx2021,DeLange2023,heckotter2023}. Our work complements efforts to prepare quantum states of light using Rydberg-mediated interactions in \cuprite~\cite{Khazali2017,Heckotter2021-2,Morin2022,Walther2023}, and opens a potential route to microwave-optical conversion in the quantum regime.

\appendix
\section{Experimental details}\label{App:experiment}
The \cuprite~sample was prepared from a natural gemstone (Tsumeb mine) using the  procedure detailed in~\cite{Lynchx2021}, and glued to a CaF$_2$ window. The light had an incident intensity of 20~\si{\micro\watt}/mm$^2$, propagated along the [111] crystallographic axis, and was linearly polarized. Finite element analysis showed that the microwave electric field was aligned in the plane of the sample and orthogonal to stripline. $\Delta \alpha$ was observed to depend very weakly on the angle between microwave and optical electric fields with a less than $10\%$ difference observed when varying the angle by $\pi/2$~\cite{PritchettThesis}. No microwave-induced change in the polarization of the transmitted light was observed within the experimental uncertainty. The etalon had a free spectral range $60.1\pm0.2$ GHz and a finesse of $44.5\pm0.7$, and was temperature-tuned. The carrier and sideband amplitudes were extracted by fitting the Lorentzian response function of the etalon to spectra like those in Fig.~\ref{fig:setup}(c).

As stated in the text, the nonlinear response is very broadband, easily covering the range 1-20 GHz. However  antenna resonances make comparing multiple frequencies challenging \cite{Gallagher2022}. We therefore fixed $\omega_\mathrm{MW}=2\pi\times 7$~GHz to give good resolution of the sidebands while remaining in the relevant range for superconducting quantum circuits.   

\section{Sideband susceptibility}\label{App:sideband_chi}

The nonlinear susceptibility at the frequency of the red and blue sidebands is

\begin{equation}
\begin{split}
    &\chi^{(3)}_{n\mathrm{P}n'l'n''\mathrm{P}}(\omega_\mathrm{L}\pm2\omega_{\mathrm{MW}}) =\\ &\frac{N_\mathrm{d}}{2\epsilon_0 \hbar^3} \frac{D^{\mathrm{VB}\rightarrow n\mathrm{P}} d^{n\mathrm{P}\rightarrow n'l'}d^{n'l'\rightarrow n''\mathrm{P}} D^{n''\mathrm{P}\rightarrow\mathrm{VB}}}{(\delta_{n\mathrm{P}}-i\Gamma_{n\mathrm{P}})(\delta_{n'l'}^{\pm\omega_\mathrm{MW}}-i\Gamma_{n'l'})(\delta^{\pm2\omega_\mathrm{MW}}_{n''\mathrm{P}}-i\Gamma_{n''\mathrm{P}})}.
\end{split}
\label{eq:chi3SB}
\end{equation}
The only difference between this and Eq.~\ref{eq:chi3carrier} is the detuning in the third term on the denominator, which is given by $\delta^{\pm2\omega_\mathrm{MW}}_{n'\mathrm{P}}=\omega_{n'\mathrm{P}} -(\omega_\mathrm{L} \pm 2\omega_\mathrm{MW})$.
In principle, it is possible to derive the sideband amplitude by using these susceptibilities as source terms. However this calculation is strongly dependent on the unknown relative phase of each term in the summation~\cite{Rogers2022}.

\section{Parameters for model}\label{App:parameters}
Here we give details of the parameters in the susceptibility model (Eq.~\ref{eq:chi3carrier}). States from $n=5$ to $17$ and $l=0$ to 2 are included in the model. 

Matrix elements for transitions between exciton states, $d$, were calculated using a spinless hydrogen-like model \cite{Walther2020} for the exciton states, with quantum defects obtained from dispersion relations for the electron and hole~\cite{Walther2018}.

The widths, energies and matrix elements $D$ for transitions from the valence band to the $n$P exciton state were extracted from experimental absorption spectra without microwaves. Values for the energy and width of the S and D states were extracted from~\cite{Rogers2022}. In the two-photon experiments only states up to $n=12$ were observed and so extrapolation was used to extend the range of $n$ beyond that available from these experiments.  The energies were extrapolated using a quantum defect model
\begin{equation}
    E_{nl} = E_\mathrm{g} - \frac{R_\mathrm{X}}{(n-\delta_{l})^2},
\end{equation}
where $R_\mathrm{X}$ is the excitonic Rydberg energy and $\delta_{l}$ is the quantum defect.

The widths of the P states were fitted using the equation 
\begin{equation}
    \Gamma_{n\mathrm{P}} = \frac{\Gamma}{n^3} + \Gamma_0,
\end{equation}
where $\Gamma_0$ is constant offset attributed to an inhomogenous broadening due to charges and defects in the material~\cite{Kruger2020}. For the S and D widths, the $\Gamma_0$ was fixed to the value fitted from the P state trend and the high $n$ states were extrapolated. 

\begin{backmatter}
\bmsection{Funding} This work was supported by the Engineering and Physical Sciences Research Council (EPSRC), United Kingdom, through research grants EP/P011470/1, EP/P012000/1, EP/X038556/1 and EP/X03853X/1.

\bmsection{Acknowledgments} We thank I. Chaplin and S. Edwards 
for preparing the samples, and V. Walther for calculating the dipole matrix elements.

\bmsection{Disclosures} The authors declare no conflicts of interest

\bmsection{Data availability} Data supporting this work is available at doi:10.15128/r27d278t10x

\end{backmatter}

\bibliography{Kerr_paper}


\end{document}